\documentclass[pre,footinbib,twocolumn,preprintnumbers,amsmath,longbibliography,amssymb]{revtex4-2}

\usepackage{amssymb}
\usepackage{amsmath}

\usepackage{color}
\usepackage[pdftex]{graphicx}

\usepackage{empheq}
\usepackage{mathtools}

\usepackage{makecell}
\usepackage{subfigure}
\usepackage{enumerate}
\usepackage{mathtools}
\usepackage{stmaryrd}
\usepackage{enumitem}
\usepackage{textcomp}
\usepackage{gensymb}

\usepackage{natbib}

\newcommand{\ex}[1]{\mathrm{e}^{#1}}

\newcommand{\rr}[0]{\boldsymbol{r}}

\newcommand{\kB}[0]{k_{\mathrm{B}}}

\newcommand{\dw}[0]{\Delta w}
\newcommand{\dmu}[0]{\Delta \mu}

\newcommand{\cC}{\mathcal{C}}

\definecolor{darkblue}{rgb}{0,0,0.6}
\definecolor{darkred}{rgb}{0.6,0,0}
\usepackage[colorlinks=true,urlcolor=darkblue,citecolor=darkblue,linkcolor=darkred]{hyperref}

\begin{document}

\title{Control of encounter kinetics by chemically active droplets}

\author{Jacques Fries}
\affiliation{Sorbonne Universit\'e, CNRS, Physico-Chimie des \'Electrolytes et Nanosyst\`emes Interfaciaux (PHENIX), 4 Place Jussieu, 75005 Paris, France}

\author{Roxanne Berthin}
\affiliation{Sorbonne Universit\'e, CNRS, Physico-Chimie des \'Electrolytes et Nanosyst\`emes Interfaciaux (PHENIX), 4 Place Jussieu, 75005 Paris, France}

\author{Marie Jardat}
\affiliation{Sorbonne Universit\'e, CNRS, Physico-Chimie des \'Electrolytes et Nanosyst\`emes Interfaciaux (PHENIX), 4 Place Jussieu, 75005 Paris, France}

\author{Pierre Illien}
\affiliation{Sorbonne Universit\'e, CNRS, Physico-Chimie des \'Electrolytes et Nanosyst\`emes Interfaciaux (PHENIX), 4 Place Jussieu, 75005 Paris, France}

\author{Vincent Dahirel}
\thanks{vincent.dahirel@sorbonne-universite.fr}
\affiliation{Sorbonne Universit\'e, CNRS, Physico-Chimie des \'Electrolytes et Nanosyst\`emes Interfaciaux (PHENIX), 4 Place Jussieu, 75005 Paris, France}

\begin{abstract}
	Biomolecular condensates play a crucial role in the spatial organization of living matter. These membrane-less organelles, resulting from liquid-liquid phase separation, operate far from thermodynamic equilibrium, with their size and stability influenced by non-equilibrium chemical reactions. While condensates are frequently considered optimized nanoreactors that enhance molecular encounters, their actual impact on reaction kinetics remains unclear due to competing effects such as diffusion hindrance, and random trapping in non-specific condensates.
	In this study, we develop a microscopic, stochastic model for chemically active droplets, incorporating reaction-driven modulation of protein interactions. Using Brownian dynamics simulations, we investigate how protein interactions and active coupling to a free energy reservoir influence phase separation, molecular transport and reaction kinetics. We demonstrate that the intensity of the chemical drive governs surface dynamics, generating fluxes that modulate bimolecular reaction rates. Comparing active emulsions to homogeneous systems, we reveal that condensates can either accelerate or decelerate molecular encounters. Our findings provide key insights into the role of biomolecular condensates as potential regulators of intracellular reaction kinetics.
\end{abstract}

\maketitle

\section{Introduction}

The nanoscale organization of soft and living materials critically influences the kinetics of 
chemical reactions between diffusing particles. Geometric constraints affect the distribution and transport of these particles~\cite{benichou2010geometry}, thus shaping the efficiency and selectivity of their reactions. Geometry-controlled kinetics have been described in many systems~\cite{minton2008,zhou2018micelles,corma1997microporous,narayanan2004shape,ferey2008hybrid,weckhuysen2019microfluidics}, and shall be prominent in biological media.
Compartmentalizing the cell into organelles allows for a fine control of the local composition in biomolecules, and therefore helps tuning the characteristics of chemical reaction networks within each compartment~\cite{hinzpeter2017optimal,li2020reprogramming,kondrat2022enzyme,diamanti2023surpassing}. 
Some of these compartments, called biocondensates, form through liquid-liquid phase separation (LLPS)~\cite{Hyman2014,Weber2019}. 
These membrane-less structures play a key role in the regulation of multiple metabolic processes~\cite{Banani2017, Shin2017}. 

Many studies suggest that biocondensates optimize the kinetics of enzymatic reactions~\cite{Banani2017,Kojima2018,Hinzpeter2019,Kuffner2020,Peeples2021,Flynn21,ranganathan2023enzymatic,Smokers2024}. 
Condensates locally increase the concentration of embedded biomolecules by several orders of magnitude~\cite{Flynn21}. 
While decreased diffusion distances increase molecular encounter rates~\cite{Smol1916,benichou2010geometry}, other physico-chemical factors may balance their benefits. Attracting interfaces~\cite{najafi2023size} and strong crowding within mesoscale compartments slow down the long time transport of biomolecules~\cite{Ellis2003,Zimmerman1993,Minton2015,Saini2023}. Single Particle Tracking~\cite{steves2024single} shows that the diffusion coefficient of proteins is significantly lowered within the condensates~\cite{smigiel2022protein,Lorenz-Ochoa23}. 
For example, an up to 500-fold
decrease of the translational diffusion coefficient was measured for the low complexity domain (LCD)
of the human fused in sarcoma (FUS) protein in condensates~\cite{murthy2019}. In addition, cells contain a very large number of condensates~\cite{Keber2024} that may act as molecular insulators. For instance, images of the cytoplasm of C. Elegans cells show up to $200$ coexisting RNA granules~\cite{Wang2014}. When the molecular meeting partners are diluted, these viscous condensates form a dense suspension of traps that potentially increase the diffusion time until the partners encounter~\cite{Flynn21,lipinski2022biomolecular}. 
Lastly, the relevance of classical theories of diffusion-limited encounter has not been established for biological condensates, as the underlying assumptions are not valid for non-equilibrium systems. 
Indeed, some  chemical reactions such as ATP hydrolysis are thought to bring free energy to the droplets, making them active~\cite{Zwicker2017, Weber2019,Guilhas2020,Zwicker2022}. These reactions within condensates are known to significantly impact phase separation~\cite{Tjhung2018,Hondele2019,Kim2019,Linsenmeier2022,Ziethen2023,Wu2024}. 

Therefore, given the many competing effects at stake (spatial confinement, molecular crowding, non-specific interactions...), the role of LLPS in controlling encounters between biomolecules is still poorly understood. In particular, it appears crucial to determine under which conditions the formation of droplets by phase separation can decrease the encounter times between biomolecules and, therefore, speed up reactions. In addition, when the droplets stem from active reactions~\cite{Soding2020,Gouveia2022}, the role of activity in the promotion of biomolecule encounters remains unclear. 

To address these issues, in this article, we resort to a microscopic and stochastic model of phase-separated systems at equilibrium or under the influence of active chemical reactions. The properties of these systems are investigated by numerical simulations~\cite{berthin2025,fries_active_2024}. Our Brownian Dynamics (BD) simulation technique~\cite{berthin2025} takes into account the effect of the system geometry, and of all the interactions between biomolecules on the transport coefficients and reaction probabilities. 
In order to investigate a large range of parameters, and to help rationalizing our results, we extract from these BD simulations the parameters of a mean-field model, which is a generalization of diffusion-limited reaction kinetics~\cite{Smol1916} to a medium divided into mesoscale compartments. This bottom-up approach helps a more systematic exploration of the role of the different parameters. In particular, it unravels how matching the residence time and the encounter time in droplets leads to optima in the mean encounter kinetics in active phase-separated media. 

\begin{figure}
	\centering
	\includegraphics[width=\columnwidth]{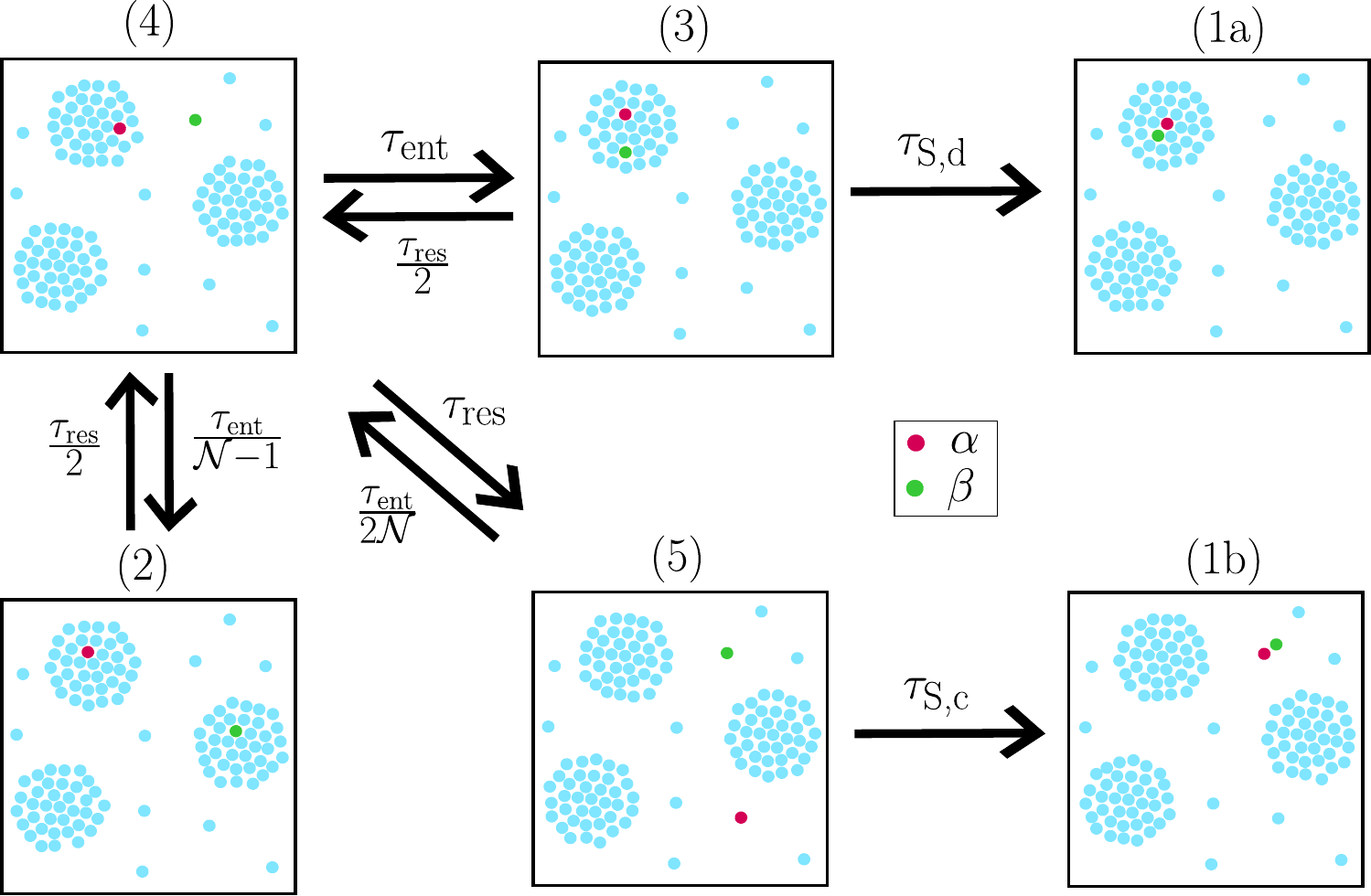}
	\caption{\textbf{Schematic representation of the 5 states of the mean field model}, for a case with $\mathcal{N}=3$ droplets. The transition times between states are indicated above each arrow. State (1) corresponds to all cases for which the particles $\alpha$ and $\beta$ have met at least once, either in a droplet (meeting represented as (1a)) or in the continuous phase (1b).}
	\label{fig:scheme}
\end{figure}

\section{Mean field model for encounter kinetics in mesostructured media}

Our main focus concerns the time of microscopic events, such as chemical reactions, that are governed by encounters between molecules. In a finite-size homogeneous medium of volume $V$, the time $\tau_{\rm S}$ needed for two particles $\alpha$ and $\beta$ to meet has first been estimated by Smoluchowski~\cite{Smol1916}:  $\tau_{\rm S}=V/(8 \pi D\sigma)$, when both particles have the same diameter $\sigma$ and diffusion coefficient $D$. In this calculation, 
the particles start from a random position and diffuse. Encounter is defined by the first moment they approach at a distance equal to their diameter.

We adapt this classical treatment to the description of encounters in a medium divided into $\mathcal{N}$ droplets of volume $v$ in a continuous phase of volume $V - \mathcal{N}v$. Our resulting mean field model describes the evolution of $p_n(t)$, $ 1 \le n \le 5$, the probability that the system is in one of the following states (see Fig.~\ref{fig:scheme}). State (1) is an absorbing state where particles $\alpha$ and $\beta$ have met at least once. In states (2)-(5), $\alpha$ and $\beta$ have never met. In state (2), $\alpha$ and $\beta$ are in two distinct droplets; in state (3), $\alpha$ and $\beta$ are in the same droplet; in state (4) $\alpha$ is in a droplet, and $\beta$ is in the continuous phase (or the opposite), and in state (5), $\alpha$ and $\beta$ are in the continuous phase. The evolution of $p_n(t)$ follows coupled first order kinetic equations, whose rates are the inverse of the following times: (i) the Smoluchowski time in a droplet, $\tau_{\rm S,d}=v/(8 \pi D_{\rm d}\sigma)$, and in the continuous phase, $\tau_{\rm S,c}=(V-\mathcal{N}v)/(8 \pi D_{\rm c}\sigma)$, where $D_{\rm d}$ (resp. $D_{\rm c}$) is the diffusion coefficient of the particles in a droplet (resp. in the continuous phase), (ii) the residence time $\tau_{\rm res}$, the average time spent by a particle in a droplet, and (iii) the entering time $\tau_{\rm ent}$, the average time it takes a particle to move from the continuous phase into a specified droplet. 
The master equations of the model read:
\begin{align}
	\frac{{\rm d}p_1}{{\rm d} t}&= p_3/\tau_{\rm S,d} + p_5/\tau_{\rm S,c} \\
	\frac{{\rm d}p_2}{{\rm d} t}&=-2p_2/\tau_{\rm res}+ (\mathcal{N}-1)p_4/\tau_{\rm ent}\\
	\frac{{\rm d}p_3}{{\rm d} t}&=-(2/\tau_{\rm res}+1/\tau_{\rm S,d})p_3+ p_4/\tau_{\rm ent}\\
	\frac{{\rm d}p_4}{{\rm d} t}&=\mathcal{N}(2p_5-p_4)/\tau_{\rm ent} + 2(p_2+p_3)/\tau_{\rm res} -p_4/\tau_{\rm res}\\
	\frac{{\rm d}p_5}{{\rm d} t}&=-(2 \mathcal{N}/ \tau_{\rm ent}+1/\tau_{\rm S,c})p_5+ p_4/\tau_{\rm res}
	\label{eq_Smoluch}
\end{align}

We numerically solve these master equations, and find that $p_1(t)$ decays exponentially with time in the long-time regime. The mean first encounter time $\mathcal{T}_e$ is defined as this characteristic decay time: $\ln(1-p_1(t)) \propto -t/\mathcal{T}_e$. We characterize the evolution of the encounter time $\mathcal{T}_e$ for parameters that are consistent with microscopic models of particles forming droplets through liquid-liquid phase separation.
These parameters are extracted from Brownian Dynamics (BD) simulations of phase separating particles. As we proceed to show, 
our mean-field model fed by these extracted parameters reproduces qualitatively well the long-time encounter dynamics predicted by the BD simulations.   

\section{Equilibrium Brownian Dynamics of phase separated systems}

\begin{figure*}
	\centering
	\includegraphics[width=0.5\linewidth]{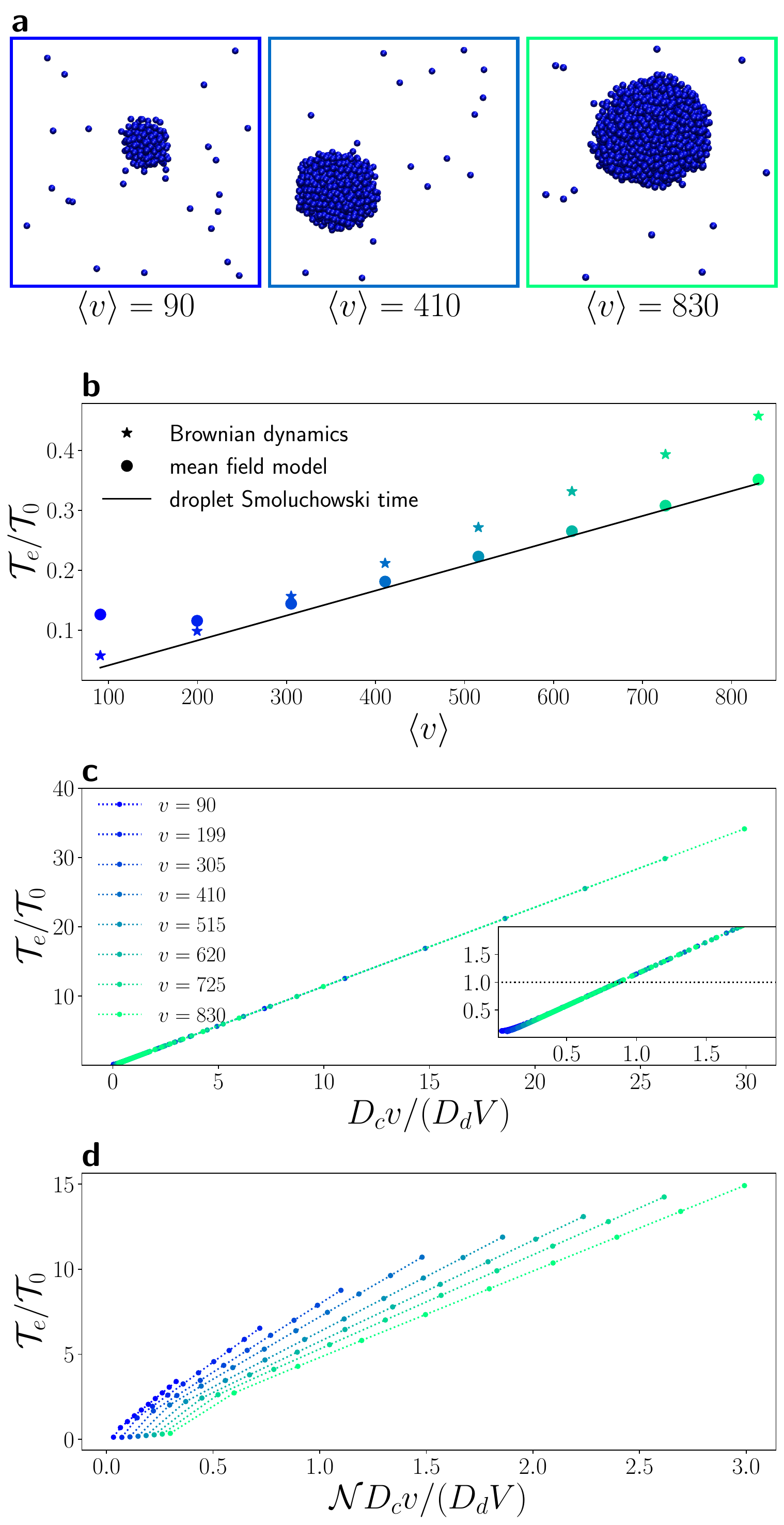}
	\caption{\textbf{(a) Snapshots from Brownian Dynamics (BD) simulations}. $B$ particles are represented as blue spheres.    
		\textbf{(b-d) Mean encounter time} $\mathcal{T}_e$ between particles divided by the reference encounter time in a homogeneous system of similar volume, $\mathcal{T}_0$. \textbf{(b) Equilibrium system with a single droplet} ($\mathcal{N}=1$). Stars show the results from BD simulations, while circles show the predictions from the mean field model whose input parameters are extracted from BD. The triangles show the Smoluchowski time in droplets, divided by  $\mathcal{T}_0$. 
		\textbf{(c) Influence of diffusion hindrance in the droplet}. Results from the mean field model for cases with a single droplet ($\mathcal{N}=1$). The master equations are solved numerically for values of $D_{\rm c}$, $\tau_{\rm res}$ and $\tau_{\rm ent}$ extracted from the BD simulations. The diffusion coefficient in the droplet $D_{\rm d}$ varies between $10^{-3 }D_{\rm c}$ (large hindrance in droplets) to $10^{-1}D_{\rm c}$ (low hindrance). The volume $v$ of the droplet  varies between $90$ and $830$ (colors). All data collapse on a single straight line. This linear behavior holds when droplets accelerate encounters ($\mathcal{T}_e < \mathcal{T}_0$ (inset)).  \textbf{(d) Influence of trapping by multiple droplets}. Results from the mean field model for many droplet systems, $1 \le \mathcal{N}\le 10$. The master equations are solved numerically for values of $D_{\rm c}$, $D_{\rm d}$, $\tau_{\rm res}$ and $\tau_{\rm ent}$ extracted from the BD simulations, and $v$ between $90$ and $830$ (colors).} 
\label{fig:equilibrium}
\end{figure*}

We choose the Lennard-Jones (LJ) fluid as a generic microscopic model of a phase separating system~\cite{Smit1991}. In the context of biocondensate modeling, a LJ particle represents a protein from a condensate. 
We simulate suspensions of $N_B$ diffusing particles called $B$, in a finite cubic volume with periodic boundary conditions. $N_B$ varies between $200$ and $1600$. $B$ particles interact with each other through a LJ potential, with an energy well of depth $\varepsilon=2 \kB T$. This ensures that the $B$ particles form droplets at the chosen density, as shown in the simulation snapshots (Fig.~\ref{fig:equilibrium}\textbf{a}). 
The positions of particles %$\{\rr_i\}$ 
obey overdamped Langevin equations, with the same bare diffusion coefficient $D$ for all particles.
At steady state, we observe that the dense phase forms a single droplet ($\mathcal{N} = 1$), whose volume is proportional to the volume fraction in LJ particles. 
As a reference, we also consider homogeneous systems ($\mathcal{N} = 0$) at similar volume fractions, where interactions between particles are purely repulsive (Weeks-Chandler-Andersen (WCA) potential~\cite{Weeks1971}).  In what follows, the distances are measured in units of the diameter $\sigma$ of LJ/WCA particles, the energies in units of $\kB T$, and time in units of $\sigma^2/D$.

We consider a pair of particles to encounter when they approach each other at a distance smaller than their diameter for the first time. 
For a given initial position of any pair of particles, we can compute the number of encounters as a function of time, and deduce the survival probability, defined as the probability that a pair has never encountered up to time $t$. This quantity is averaged over $5$ realizations starting from different configurations representative of the steady state. The resulting probability distribution displays an exponential regime at long time, which defines the mean encounter time $\mathcal{T}_e$ consistently with the definition of $\mathcal{T}_e$ in our mean field model.  

Fig.~\ref{fig:equilibrium}\textbf{b} shows the mean encounter time between particles for different values of the droplet volume $v$. 
With the parameters extracted from the BD trajectories, we find a very good qualitative agreement between the mean field model and simulation data for the encounter time. Note that none of the model parameters is adjusted to make this comparison. Both treatments predict a linear increase of $\mathcal{T}_e$ with $v$ for all but the smallest droplet ($v > 200$). It indicates that surface terms may be negligible, and most encounters occur inside the droplets. We indeed find that $\mathcal{T}_e$ is very close to the Smoluchowski time in a droplet $\tau_{\rm S,d}=v/(8 \pi D_{\rm d}\sigma)$. 
In all the investigated systems, 
the mean encounter time is smaller than $\mathcal{T}_0$, the encounter time for the reference homogeneous system. 
The particles are confined inside the droplet for a sufficient time to ensure a high probability of encounter ($\tau_{\rm S,d} \ll \tau_{\rm res}$, not shown). This scenario corresponds to the usual qualitative description of condensates as nanoreactors~\cite{Banani2017, Shin2017}. The rate enhancement depends on the size of the droplet: encounters are faster within smaller compartments. 

Nevertheless, above a critical droplet volume $v_c$, the 
Smoluchowski time in the droplet $\tau_{\rm S,d}$ exceeds that of the reference homogeneous system $\mathcal{T}_0$. The volume $v_c$ depends on the diffusivity hindrance in the droplet. In our simulations, the diffusion coefficient is roughly $16$ times smaller in the droplet than in the continuous phase. In real biocondensates, since proteins are partly disordered and may be mixed with RNA~\cite{Shin2017}, the viscosity inside droplets gets very large. Diffusion is lowered in condensates up to a factor $500$~\cite{murthy2019,smigiel2022protein,Lorenz-Ochoa23,Mazzocca2023}. For instance, in DNA repair condensates, a $50$ fold reduction of the diffusion coefficients of Rad52 proteins has been reported~\cite{Min-Hattab2021}. With our mean field model, we can systematically establish the influence of the diffusion coefficient of particles in the droplet $D_{\rm d}$ on the mean encounter time $\mathcal{T}_e$. 
By varying both $D_{\rm d}$ and the volume of the droplet $v$, we find that all the data collapse on the same line of equation $\mathcal{T}_e/\mathcal{T}_0 = 1.14 D_{\rm c}v/(D_{\rm d}V)$ (Fig.~\ref{fig:equilibrium}\textbf{c}). We can plug in this equation some typical numbers from biological systems. Some experiments have shown that roughly $20\%$ of all biomolecules are part of droplets~\cite{Keber2024}.
If we consider that $30\%$ of the cell volume is occupied by biomolecules~\cite{Minton2015}, then, if there is a single droplet, we get $v/V \approx 0.06$. In this scenario, phase separation starts becoming unfavorable ($\mathcal{T}_e > \mathcal{T}_0$ ) for $D_{\rm d} < 0.05 D_{\rm c}$.

So far, we have neglected the coexistence of many droplets in biological media~\cite{Wang2014,Shin2017}. We now use our mean field model to study the impact of increasing the number of droplets $\mathcal{N}$, while keeping the remaining parameters from the equilibrium BD simulations with $\mathcal{N}=1$. 
Fig.~\ref{fig:equilibrium}-\textbf{d} shows the dramatic impact of multi-droplet geometries: with the mean field model, the encounter time $\mathcal{T}_e$ linearly grows with $\mathcal{N}$. 
This suggests that, while two particles in the same droplet rapidly meet, these particles may also spend a long time trapped in distinct droplets. The curve collapse for $\mathcal{N} > 1$ is not as good as with single droplet systems. Nevertheless, %within a $50 \%$ margin of error, 
the evolution is qualitatively well captured by a linear scaling of $\mathcal{T}_e(\mathcal{N})$, of equation $\mathcal{T}_e/\mathcal{T}_0 = \alpha \mathcal{N} D_{\rm c}v/(D_{\rm d}V)$, with $\alpha$ a scaling factor of the order of $5$ to $10$. 

We apply this formula to two choices of typical biological numbers. In a first scenario, we consider a single type of condensate, the RNA granules of \emph{C. Elegans} cells. There are typically $\mathcal{N} = 10^2$ RNA granules per cell~\cite{Wang2014}. We estimate the typical diameter of a granule to be $200$ nm, and that of a cell to be $10$ $\micro m$, so $v/V \simeq 10^{-5}$. We assume a $10^2$-fold increase of viscosity in the condensate ($D_{\rm c}/D_{\rm d} = 10^2$). This leads to $\mathcal{T}_e/\mathcal{T}_0 = 0.25$, a $4$-fold faster encounter in the presence of condensates. In a second scenario, we consider all the biocondensates of a cell as potential traps. The parameters are similar, except the number of droplets $\mathcal{N}$, which now satisfies  $\mathcal{N}v = 0.06 V$ to keep a fraction of the volume occupied by droplets equal to $6\%$. This leads to $\mathcal{N}=6 \times10^{3}$, and thus  $\mathcal{T}_e/\mathcal{T}_0 = 15$. In this case, there is a $15$-fold slower encounter in the presence of condensates. Our results suggest that the trapping of particles in condensates may significantly increase the time to find a molecular partner.   

Nevertheless, this effect of $\mathcal{N}$ may not be relevant to many biological systems. Indeed, it is now established that the coexistence of several droplets is related to the arrest of Oswald ripening by chemical reactions~\cite{berthin2025,fries_active_2024}. The model parameters we deduced from equilibrium simulations may not be physically relevant for such non-equilibrium systems. We thus explore the encounter kinetics in active emulsions, using a Brownian Dynamics method that we designed in a previous work~\cite{berthin2025}.

\section{Brownian model of active emulsions}

In our model of active emulsions, the $B$ particles exist in another state $A$ that is not prone to aggregation.
In a biological context, this loss of condensation propensity would come from a modulation of effective interactions between proteins induced by chemical modifications~\cite{Snead2019}.
We also add chemically inert particles $C$ to keep the total volume fraction $\phi$ constant ($\phi = 0.1$).
$A$ and $C$ particles interact with any particle with a purely repulsive WCA potential~\cite{Weeks1971}. 

The $A    \xrightleftharpoons[]{}   B$ reaction is described by a random telegraph process~\cite{Gardiner1985}. 
The reaction probabilities follow two kinds of law, depending on the local density in particles. In dilute regions (the continuous phase), the probabilities meet the detailed balance condition: $P_{A    \to  B} / P_{B    \to   A} = \exp(-\Delta E)$, where $\Delta E$ is the variation of energy associated with the conversion of a $A$ particle into $B$. 
A key parameter is thus the difference in internal energy of $A$ and $B$, called $\Delta w$. Activity emerges from a coupling to an external source of chemical free energy. This chemostat operates in dense regions (droplets), where active reactions dominate, and break detailed balance: $P_{A    \to  B} / P_{B    \to   A} = \exp(-\Delta E + \Delta \mu)$, where $\Delta \mu$ is a chemical drive. In biological systems, the usual energy source comes from the conversion of ATP in ADP. In this context, $\Delta \mu$ corresponds to the reaction free energy of ATP hydrolysis.  

In what follows, all quantities are averaged over time and over independent realizations at steady state. Here, steady state means that the mean quantities do not vary when the time window for averaging is extended.

\section{Encounters in active systems are not fully controlled by the structure of the system}

\begin{figure*}
\centering
\includegraphics[width=\linewidth]{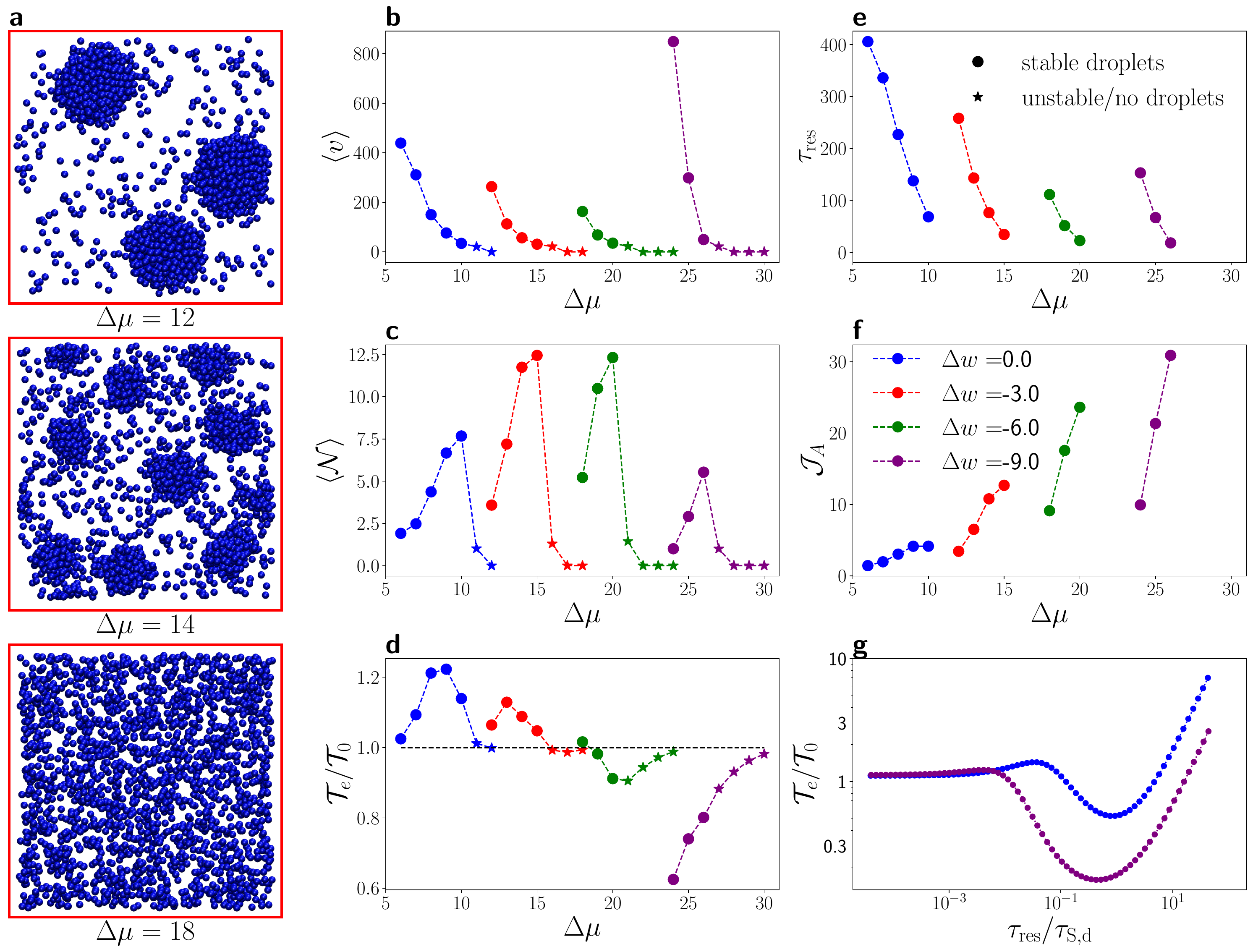}
\caption{\textbf{(a) Snapshots from reactive Brownian Dynamics simulations}, for $\dw = -3.0$, and three values of $\dmu$. $A$ and $B$ particles are both represented as blue spheres. For $\dmu=18$, droplets are unstable, and the fluid is homogeneous. 
\textbf{(b-f) Properties of droplets computed at stationary state in BD simulations}, as functions of the chemical drive $\dmu$, for several values of the internal energy difference $\dw$. In each plot, unstable systems are represented with a star symbol (see text). 
\textbf{(b) Mean volume of individual droplets}, defined as the number of particles in a droplet, times the volume of a particle.    
\textbf{(c) Mean number of droplets}. \textbf{(d) Mean encounter time between particles}, defined as the characteristic time associated with a long time exponential fit of the survival probability of encounter between two particles (see methods). \textbf{(e) Net rate of $A$ escape} from droplets  \textbf{(f) Residence time }in droplets $\tau_{\rm res}$. \textbf{(g) Influence of the ratio of the residence and encounter time scales in droplets}, $\tau_{\rm res}/\tau_{\rm S,d}$, on the mean encounter time $\mathcal{T}_e$ between pair of droplet clients divided by the reference encounter time in a homogeneous system of similar volume. These results are obtained with the mean field model, which is solved numerically for values of $v$, $\mathcal{N}$, $D_{\rm c}$, $D_{\rm d}$ and $\tau_{\rm ent}$ extracted from the BD simulations (blue or violet colors corresponding to two sets of simulations). $\tau_{\rm S,d}$ is fixed in the calculations, but $\tau_{\rm res}$ varies.
}
\label{fig:average_encounter_time_dp_volume_number}
\end{figure*}

As described in a previous study~\cite{berthin2025}, the usual mechanisms at play in liquid-liquid phase separation (spinodal decomposition, Ostwald ripening, coalescence) emerge from the attractive interactions between $B$ particles, while chemical reactions limit droplet growth and lead to droplet size selection. In Fig. \ref{fig:average_encounter_time_dp_volume_number}-\textbf{a}, snapshots of the simulations show two systems with stable droplets of distinct selected sizes, and a system where the chemical reactions completely kill phase separation.

A cluster analysis allows us to compute the volume $v$ of droplets in the simulation box. The mean volume $\langle v \rangle$ and the mean number of droplets $\langle \mathcal{N} \rangle$ depend both on $\Delta w$, the difference of internal energy between $A$ and $B$, and on the chemical drive $\dmu$ (Fig. \ref{fig:average_encounter_time_dp_volume_number}-\textbf{b} and Fig. \ref{fig:average_encounter_time_dp_volume_number}-\textbf{c}).  $\langle v \rangle$  varies by two orders of magnitude in the investigated systems. At constant  $\dw$, $\langle v \rangle$ decreases when the chemical drive $\dmu$ increases, so that droplets get unstable over a threshold value, called $\dmu_{\rm crit}$. In this limit ($\dmu > \dmu_{\rm crit}$), the mean encounter time $\mathcal{T}_e$ converges towards the value $\mathcal{T}_0$ of the reference homogeneous system (Fig.~\ref{fig:average_encounter_time_dp_volume_number}-\textbf{d}). 
For $\dmu < \dmu_{\rm crit}$, the encounters are either slowed down or sped up compared to the homogeneous case. Moreover, we find that this change of regime is not purely dictated by the steady state geometry of the discontinuous phase: 
in particular, the encounter dynamics is slower than in the homogeneous reference for ($\dw=0,\ \dmu=7$), but faster than in the reference for ($\dw=-9,\ \dmu=25$), although there is the same average number and volume of droplets in both situations. 

A similar geometry of the phase separated system -here described by the size and number of droplets- can lead to different mean encounter times if the time to enter or to exit a droplet varies. In systems for which chemical and mechanical equilibrium is satisfied in all points of space, such variation is not physically possible unless the material itself changes (which could be modeled by modified values of the interaction parameters).
Indeed, in a hypothetical system with a fixed geometry of the discontinuous phase, with fixed energy barriers to pass from the discontinuous to the continuous phase, the evolution of the system is fully determined by diffusion equations and barrier crossing probabilities. 
This is no longer true for active systems. 
The analysis of our simulations shows that the residence time $\tau_{\rm res}$ strongly decreases with $\dmu$ (Fig.~\ref{fig:average_encounter_time_dp_volume_number}-\textbf{e}).
In order to better understand this dependence, we now analyze the flows across the interface between the droplets and the continuous phase.

\section{Time scale matching leads to encounter optimization}

In active systems, $B$ particles constantly enter the droplets at steady state, and convert into $A$ ones that leave the droplets~\cite{Zwicker2022}. 
A mass flow arises from the conversion of chemical energy into mechanical energy~\cite{Decayeux2021a,Gouveia2022}. 
By tracking individual droplets, we compute the mean steady-state flow of $A$ particles exiting a droplet, $ \mathcal{J}_A$.
At fixed $\dw$, we find that the amplitude of mass flows increases with $\dmu$ (Fig.~\ref{fig:average_encounter_time_dp_volume_number}-\textbf{f}). This is also observed in deterministic diffusion-reaction models (see supp.mat. Fig S3). This suggests that the residence time $\tau_{\rm res}$ of specific proteins in active biocondensates is directly controlled by the chemical drive $\dmu$ associated with chemical modifications of these proteins. 

Our mean field model allows to study the influence of the residence time $\tau_{\rm res}$ while keeping all the other parameters fixed. In contrast with the equilibrium cases, there is no clear scaling between the parameters of the model and the resulting mean encounter time (Fig.~\ref{fig:average_encounter_time_dp_volume_number}-\textbf{g}). 
The behavior is much richer in active systems, with non-monotonic variations and the presence of local extrema of the encounter time. 
There is an optimal mean encounter time for particles in active emulsions when the time to encounter in a droplet $\tau_{\rm S,d}$ and the time to exit the droplet $\tau_{\rm res}$ are similar.
This result can be rationalized as follows. If a particle searches for a partner and both are in the same droplet, in average the partners need a time $\tau_{\rm S,d}$ to encounter, but do not need to stay longer in the droplet. 
In case they stay longer ($\tau_{\rm res}>\tau_{\rm S,d}$), any increase of the residence time $\tau_{\rm res}$ only adds to the time lost when the particles are confined in distinct droplets, thus increasing $\mathcal{T}_e$.  
In the limit $\tau_{\rm res} \gg \tau_{\rm S,d}$, the particles are trapped within distinct droplets and the time to meet diverges. 
In the limit $\tau_{\rm res} \ll \tau_{\rm S,d}$, the particles are not confined in droplets, and phase separation does not make any difference. 
Interestingly, there is a local maximum of the encounter time for low values of $\tau_{\rm res}/\tau_{\rm S,d}$, when droplets slightly capture the particles and slow down their diffusion, but do not keep them long enough to promote reactions. 
We checked that these results hold for different volumes of the droplets.  

To further validate these findings, we modify the model simulated with Brownian Dynamics in order to independently vary the Smoluchowski and residence times in droplets. We now distinguish two types of proteins. $A$, $B$ proteins assemble to create condensates as before ('scaffold' proteins). Additional $A'$,$B'$ proteins can reside in the condensate ('client' proteins). 
Clients $A'$, $B'$ proteins interact in the same way as the scaffold $A$, $B$ proteins. They also inter-convert through passive and active reactions. These new reactions are associated with the difference of internal energies of $A'$ and $B'$, $\dw'$, and a chemical drive $\dmu'$, distinct from $\dw$ and $\dmu$ associated with the conversion of scaffold proteins. %SUPP The simulation box contains $2000$ $A$ and $B$ particles and $300$ $A'$ and $B'$ particles. 
The parameters are chosen in such a way that scaffold proteins maintain the structure of droplets for a sufficiently long time, while client proteins are more prone to enter and leave the droplets. In other words, the geometric factors (volume and number of droplets) are mainly selected by $\dw$ and $\dmu$, while fluxes of clients at the surface of droplets are mainly controlled by $\dw'$ and $\dmu'$. This is a usual terminology for proteins within biocondensates, which are multicomponent droplets~\cite{SanchezBurgos2022} where the more concentrated scaffolds solvate more diluted clients. 

\begin{figure}
\centering
\includegraphics[width=\linewidth]{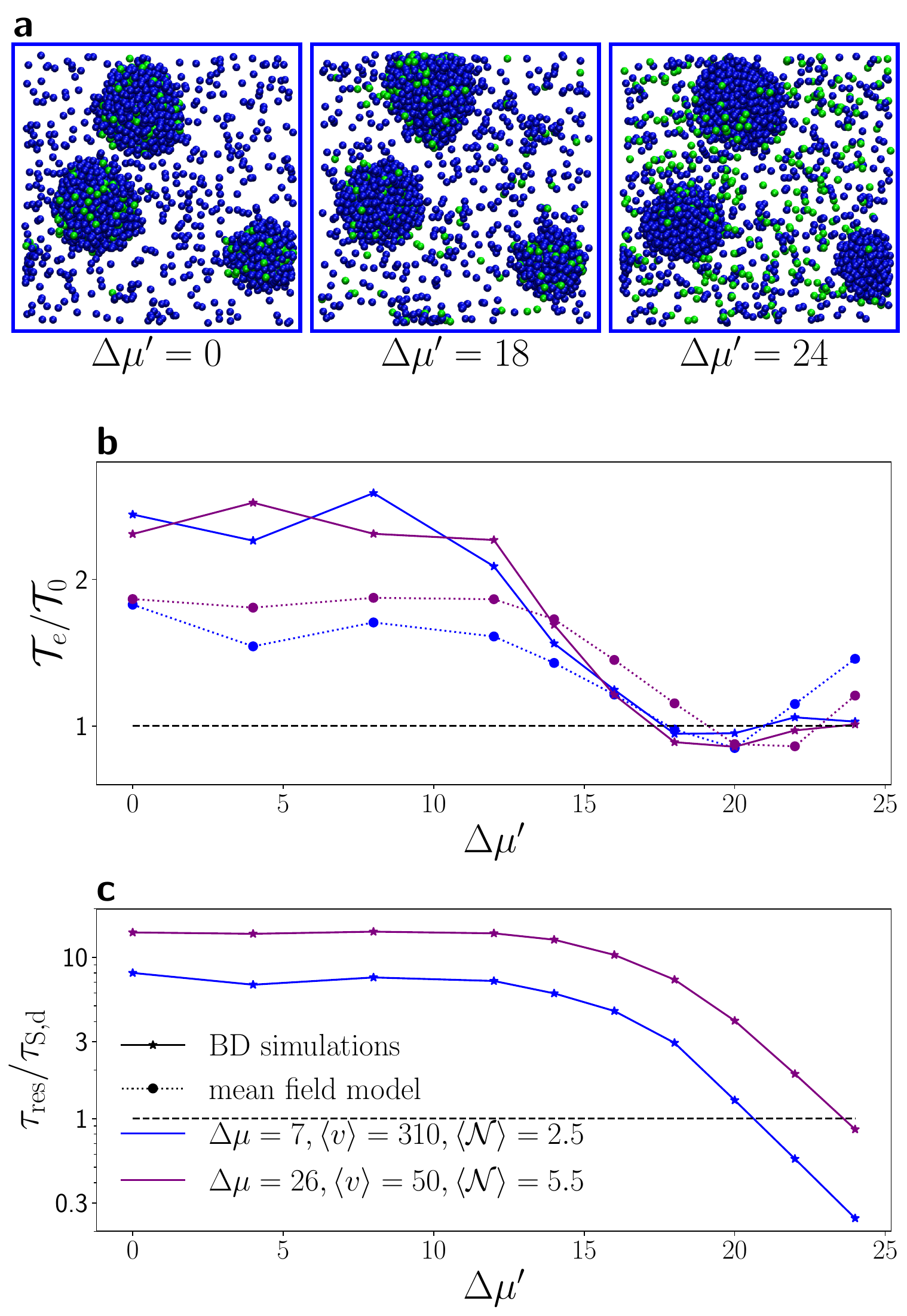}
\caption{\textbf{(a) Snapshots from reactive Brownian Dynamics simulations}, for $\dmu = 7$, and three values of $\dmu'$. Scaffold particles are represented as blue spheres, and client particles as green spheres. For $\dmu=18$, droplets are unstable, and the fluid is homogeneous. 
\textbf{(b) Mean encounter time} $\mathcal{T}_e$ between pair of droplet clients divided by the reference encounter time in a homogeneous system of similar volume, $\mathcal{T}_0$. Clients are either in $A'$ or $B'$. The internal difference of energy between clients is $\dw'=-6$. Stars show the results from Brownian Dynamics simulations, while circles show the predictions from the mean field model. 
\textbf{(c) Ratio of the mean residence time in a droplet $\tau_{\rm res}$ over the mean encounter time in a droplet $\tau_{\rm S,d}$} for the same systems. Matching of both time scales is highlighted with the line $\tau_{\rm res}/\tau_{\rm S,d} = 1$.}
\label{fig:encounter_time_trapped_enhenced}
\end{figure}

Fig.~\ref{fig:encounter_time_trapped_enhenced}-\textbf{a} show snapshots of simulations where the clients are either trapped in droplets ($\dmu'=0$), expelled from droplets ($\dmu'=24$) or in an intermediate situation. 
Fig.~\ref{fig:encounter_time_trapped_enhenced}-\textbf{b} shows the mean encounter time between clients for different values of the client chemical drive $\dmu'$. 
We find a good qualitative agreement of simulation data with the prediction from the mean field model, with no fitting parameter. 
For $\dmu'<12$, the mean encounter time of clients is almost twice larger than that of the reference homogeneous system, and $10$ times larger than that of the system at equilibrium. The active reaction does not strongly impact the transitions from $B'$ to $A'$ in this regime. 
For $\dmu'> 12$, the mean encounter time decreases with $\dmu'$. This is an expected consequence of the decrease of residence time in droplets  with the increase of chemically driven flows. There is a range of parameters for which encounters are faster than in the homogeneous system, when the residence time and encounter time within a droplet match (Fig.~\ref{fig:encounter_time_trapped_enhenced}-\textbf{c}). This effect is quantitatively similar in the mean field model and in simulations, and confirms that activity can overcome the trapping effects of multiple attracting droplets.

\section{Discussion}

The rational of nanoreactor design lies on very simple idea: 
by confining particles in a small volume, their concentration increases, which speeds up reactions. 
In biological cells, the geometric division of space into droplets suggests such design principle. In equilibrium simulations of a phase separating system with explicit particles, we could obtain an increase of encounter rates with the formation of a droplet, and quantify the influence of the physically relevant parameters, size and diffusion coefficients.  
In contrast to this expected behavior, we have also found many cases for which droplets slow down encounters. In general, this happens when several droplets coexist, and trap protein partners in distinct compartments. 
We could model this phenomenon with a mean-field model, which extends Smoluchowski diffusion-limited reaction kinetics to a system separated in distinct compartments. 
This property of multi-droplet media may be used as a tool to repress specific biomolecular interactions and inhibit parasite reactions within biochemical networks~\cite{alon2019introduction}. 

In cases where the phase separation resulting in the formation of condensates is regulated by chemical reactions, 
the active pathway controls not only the size and shape of the droplets, but also the time during which they persist and keep scaffold or client particles in close vicinity. We have shown that, in the presence of multiple active droplets, the mean encounter time between scaffold proteins forming the droplet strongly depends on the fluxes of proteins entering and exiting the droplets. These fluxes are controlled by the amplitude of the chemical drive that maintains the droplets far from equilibrium. 
Variations of $\dmu$ lead to three distinct kinetic regimes, as identified by both Brownian Dynamics simulations and our mean field model:

\begin{itemize}
\item When the residence time $\tau_{\rm res}$ in droplets is significantly larger than the time it takes for a particle to meet another in the droplet volume ($\tau_{\rm S,d}$), then the strong confinement inside droplets limits the encounter speed, since the exploration of all droplets takes a long time. 
\item Conversely, when $\tau_{\rm S,d}$ is larger than $\tau_{\rm res}$, droplets do not sequester the particles during a sufficient time to ensure a significant encounter probability. In such situation, droplets do not make any difference compared to homogeneous situations. 
\item Matching the residence time to the encounter time in droplets leads to a maximum of the kinetic rate of encounter in the active phase separated system. 
\end{itemize}

These results open new perspectives in systems biology. 
By tuning the residence time in different types of condensates with chemical (post-translational) modifications, the cell may control the kinetics of encounters between any pair of proteins. 
This control mechanism is clearly distinct from classical atomic scale models of protein-protein interactions, since it relies on mesoscopic parameters, such as the diffusion coefficients, the solubility in each phase and the free energy associated with chemical modifications (such as the phosphorylation of a protein coupled to ATP hydrolysis). 
Simple generic mechanisms implying condensates can lead to a sharp decrease or increase in the rate of reactions between two proteins. 
As a result, biocondensates could be used as agents of repression or activation of specific network elements.

\section*{Acknowledgments}

The authors thank Jean-François Joanny for discussions.

\section*{Appendices}

\subsection*{Brownian dynamics (BD) simulations}

To perform Brownian dynamics simulations, we use the LAMMPS computational package~\cite{Thompson2022}. 
We assume that the positions of particles $\rr_1,\dots,\rr_N$ obey overdamped Langevin equations. 
In the various models we simulate, there are up to $5$ species types, $A$, $B$, $A'$, $B'$, and $C$. We denote by $S_n(t)\in\{A,B,A',B',C\}$ the species of particle $n$ at time $t$. 
All the particles have the same bare diffusion coefficient $D$.
The evolution equations for the positions of particles are integrated thanks to a forward Euler-Maruyama scheme~\cite{Allen1987}:
\begin{equation}
\rr_n(t+\delta t) =\rr_n(t) + \sqrt{2D \delta t}  \boldsymbol{\xi} -\delta t\frac{D}{\kB T} \sum_{m\neq n}\nabla U_{S_n, S_m}  (r_{mn}).
\label{overdampedLangevin_disc}
\end{equation}
$\boldsymbol{\xi}$ is a random vector drawn from a normal distribution of mean $0$ and unit variance $\langle \eta_{n,i}(t)\eta_{m,j}(t') \rangle = \delta_{ij}\delta_{nm}\delta(t-t')$.

The pair interaction between two particles $m$ and $n$, denoted by $U_{S_n S_m} (r_{mn})$, depends on their species and on their relative distance $r_{mn}=|\rr_m-\rr_n|$. 
The $B$ and $B'$ particles interact with each other through a truncated and shifted Lennard-Jones (LJ) potential. 
The truncation distance is $r_c=2.5\sigma$. The potential reads 
\begin{equation}
U_{\{ B,B'\},\{ B,B'\}}(r)=[U_{\varepsilon}(r)-U_{\varepsilon}(r_c)]\theta(r_c-r)
\end{equation}

where $U_{\varepsilon}(r)$ is the standard LJ potential
\begin{equation}
U_{\varepsilon}(r)=4\varepsilon\left[\left(\frac{\sigma}{r}\right)^{12} -   \left(\frac{\sigma}{r}\right)^6 \right] 
\end{equation}

and $\theta(r)$ denotes the Heaviside function. 
The other pair interactions are modeled by a purely repulsive Weeks-Chandler-Andersen (WCA) potential~\cite{Weeks1971}, which is a  Lennard-Jones potential truncated and shifted at $r=2^{1/6}\sigma$:
\begin{equation}
\left.
\begin{array}{ll}
	U_{\{A,A',C\},\{ A,A',B,B',C\}} \\
	U_{\{B,B'\},\{A,A',C\}}
\end{array}
\right \}=[U_{\varepsilon'}(r)+\varepsilon'] \theta(2^{1/6}\sigma-r) 
\end{equation}

The energy parameters of the interaction potentials are $\varepsilon'=1 \kB T$ and $\varepsilon=2 \kB T$.  

\subsection*{Stochastic reactions}

$\text{A}    \xrightleftharpoons[]{}   \text{B}$ and $\text{A'}    \xrightleftharpoons[]{}   \text{B'}$ reactions are described by a random telegraph process~\cite{Gardiner1985}. 
A transition between two configurations $\mathcal{C}$ and $\mathcal{C}'$ is associated with a rate $k_{\mathcal{C},\mathcal{C}'}$. 
The probability of a transition from the configuration $\mathcal{C}$ at time  $t$ to $\mathcal{C}'$ at time $t + \delta t$ reads $ P(\mathcal{C}', t+\delta t |\mathcal{C}, t) = k_{\mathcal{C},\mathcal{C}'} \delta t$. 
The rates $k_{\mathcal{C},\mathcal{C}'}$ are the sum of passive and active contributions:  $k_{\mathcal{C},\mathcal{C}'} = k^{\text{p}}_{\mathcal{C},\mathcal{C}'}+k^{\text{a}}_{\mathcal{C},\mathcal{C}'}$.
To compute these rates, 
we assume that 
\begin{align}
k^{\text{p}}_{\cC',\cC} &= k_0 (1-\phi_\text{loc}/\phi_\text{max}) \ex{-\frac{\beta}{2}[E(\mathcal{C})-E(\mathcal{C}')]}, \\
k^{\text{a}}_{\cC',\cC} &= k_0 (\phi_\text{loc}/\phi_\text{max}) \ex{-\frac{\beta}{2}[E(\mathcal{C})-E(\mathcal{C}')+\kappa_{\cC',\cC}\Delta \mu]},
\end{align}

where $\phi_\text{loc}$ is the local density of all Brownian particles around the particle whose species can change between configuration $\mathcal{C}$ and $\mathcal{C'}$, and where $\phi_\text{max}$ is the maximum  volume fraction of the mixture and is approximated to the maximum packing fraction in 3 dimensions: $\phi_\text{max}\simeq 0.74$. We fix  $k_0=10^{-2}$ in all the simulations.

For each reacting particle, and at each time step, we get the number $N_\text{loc}$ of particles that are located at a distance smaller than $2.5\sigma$ (i.e. the cutoff of the interaction potential).  The local volume fraction $\phi_\text{loc}$, used to compute the reaction rates $k_{\mathcal{C}',\mathcal{C}}^\text{a,p}$ is computed as $N_\text{loc}/V_\text{loc}$, where $V_\text{loc}$ is the average of the volumes of two spheres of radius $2.5\sigma$ and $3\sigma$~\cite{berthin2025}. We then compute the difference of energy of the configurations before and after the interconversion. The reactive transition is accepted with the probabilities specified above. For computational efficiency, the species interconversion are evaluated each $10 \delta t$ or $100 \delta t$, depending on the value of the chemical drive.

\subsection*{System size for equilibrium BD simulations}

The systems contain $8000$ particles in a cubic box of length $34.729$ with periodic boundary conditions. The total volume fraction of particles is $0.1$.
The number of species $B$ particles is varied between $N_B^0=200$ and $N_B^0=1600$, completed by particles of species $C$. The particles are initially located on a face-centered cubic lattice. This initial configuration is equilibrated for $10^5$ timesteps. For each system, 5 independent realizations are performed with different seeds.

\subsection*{System size for reactive BD simulations}

The systems contain $8000$ particles in a cubic box of length $34.729$ with periodic boundary conditions. The total volume fraction of particles is $0.1$. The particles are initially located on a face-centered cubic lattice. The initial number of each species is $N_A^0=400$, $N_B^0=1600$ and $N_C^0=6000$. This initial configuration is equilibrated for $10^5$ timesteps without the attractive part of the Lennard-Jones interaction between $B$ species. Then the passive and active reactions for scaffold proteins $A$ and $B$ are switched on, and the attractive part of the Lennard-Jones interaction between $B$ species is implemented. The timestep $\delta t$ is  equal to $2\cdot 10^{-4}$.

In order to perform simulation with $A'/B'$ clients, we first run simulation without clients until the systems reach steady state~\cite{berthin2025}. Then, $300$ crowders $C$ are randomly transformed in $B'$ and the passive and active reactions for clients $A'/B'$ are switched on.

For each set of parameters with only scaffold particles $A$ and $B$ and crowders $C$, 10 independent realizations are performed with different seeds. For the simulations with clients $A'$ and $B'$, 20 independent realizations are performed for each system.

\subsection*{Cluster analysis of droplet volume and number}

To identify the droplets at each timestep, a cluster analysis of the trajectories is performed. Every  $ n\ \delta t$, the distances between the particles forming the droplets $B$ (and $B'$ in the presence of clients) are computed. Two particles are assumed to be in the same droplet if they are at a distance $d \le 1.5\sigma$~\cite{berthin2025}. The volume $v$ of a droplet is assumed to be the sum of the volume of each one of the droplet's particles.

By tracking the droplets over time~\cite{berthin2025}, we can compute the average derivative of the volume with respect to time as a function of the volume (referred to as the phase portrait of the system). We are interested in stable droplets, which corresponds to droplets bigger than the nucleation volume, the latter corresponding to the unstable fixed point of the phase portrait. Thus, only droplets made of at least 35 particles were considered.

The mean values of the number of particles in the simulation box $\langle \mathcal{N} \rangle$ and the mean volume of a droplet $\langle v \rangle$ are averaged over over time and over independent realizations.

\subsection*{Mean encounter time between Brownian particles}

In our mean field model, the solutions of the system of 5 linear differential equations are linear combinations of exponential functions $\exp{\lambda_i t}$, where we have introduced the 5 negative eigenvalues $\lambda_i$ of the system. The characteristic times of the system are defined as $\mathcal{T}_i = -1/\lambda_i$. The encounter time $\mathcal{T}_e$ is identified with the largest characteristic time.  

The encounter time can also be computed with Brownian Dynamics simulations, from the distribution of the first encounter time between pairs of scaffold or client particles.
We chose to compute such quantities for particles $i$ and $j$ regardless of their species $S_{i}(t)$ and $S_{j}(t)$ (particle $i$ in state A may encounter particle $j$ in state B, for instance). 
The time is defined by an initial state and a target state. Initial states are sampled over the configurations of the system at steady state, where particles $i$ and $j$ may be anywhere in the simulation box. 
The target 'encounter state' is reached when the distance between the two particles is lower than their diameter for the first time. The duration between these initial and target states is called the first encounter time $t_e$. 

More precisely, at each timestep $\delta t$, the distance $d$ between each neighboring particle is computed. When $d \le \sigma$ after some time $t$, we update the fraction of pairs that have encountered before $t$. This fraction is the ratio of the number of pairs of particles having encountered at least once over the total number of pairs. The time evolution of this ratio yield the cumulative distribution function of encounter $F(t)$. 

In supp. mat. Fig S2, we represent the survival probability $S(t)=1-F$.
In order to deduce an encounter time from the analysis of $S(t)$, we check that there exist a time regime where $S(t)$ behaves as $\exp(-t/\mathcal{T}_e)$. 
We find a mono-exponential behavior at long time in all cases, and thus define a mean encounter time $\mathcal{T}_e$ as the characteristic time associated with this long time exponential decay of $S$. 

\subsection*{Net exit flow of A particles}

At steady state, the average composition of droplets is constant. In order to evaluate the flux of matter across the interface of the droplets, we can thus instead compute the quantity of material that as been converted through chemical reactions. 

To compute the net reaction flux of particles inside droplets, the configurations are saved each time a reaction happens. For each of these reactive configurations, we perform a cluster analysis. For each new $A$ particle in the droplet, we check whether this particle was in this droplet in the previous configuration. In that case, we consider the  $B \to A $ reaction has happened inside the droplet. Similarly, for each newly transformed $B$ particle, we check whether it is in a droplet in the new configuration. In that case, we consider the $A \to B$ reaction has happened inside the droplet. The net number of reacted particles inside droplets is the difference between the number of reactions $B \to A $ and $A \to B$. In stationary state, it is equal to the flux of $A$ particles leaving the droplet, $\mathcal{J}_A$. 

\subsection*{Diffusion coefficients of particles in a droplet}

To compute the diffusion coefficient of A/B particles in the continuous phase, we measure the mean square displacement (MSD) for a system with $8000$ WCA particles without reaction. In this phase, we have also checked that WCA and LJ particles have very similar diffusivities. We find $D_{\rm c} \approx 0.8$. 

We compute the MSD of particles in droplets within equilibrium BD simulations with various numbers of $B$ particles (LJ particles). 
The computation of the MSD in the droplets is limited to shorter time scales, as particles may enter and exit a droplet during the course of the simulations.  
Therefore, instead of computing MSD as a function of time $t$, we compute the averaged MSD of a given particle $i$ for subparts of the trajectories of duration $\Delta t$, where the particle $i$ does not leave the droplet. We choose $3$ durations $\Delta t= 1,2,4$ (see Fig. S2), and deduce the diffusion coefficient of $B$ particles in the droplet from the average of the computed MSDs. 

We observed a systematic influence of droplet size on the diffusion coefficient  
To account for this effect, we distinguish two populations of droplet particles: particles in the core of the droplet and particles at the surface of the droplet (interface with the continuous phase). To get the diffusion coefficient in the core of the droplet, we assume that $D=D_{\rm core} \rho_{\rm core} + D_{\rm surf} \rho_{\rm surf}$, where $\rho_{\rm core}$ and  $\rho_{\rm surf}$ are the volume fractions of particles in the core and at the surface, respectively. We find $D_{\rm core}=0.05$ and $D_{\rm surf}=0.28$. For the mean field model, we consider that $D_{\rm d}=D_{\rm core}$.

\subsection*{Residence times}

To compute the mean residence time of particles inside droplets, $\tau_{\rm res}$, the configurations are saved each $10^3 \delta t$. For each configuration, we perform a cluster analysis to assign particles to specific droplets. We measure the time spent inside and outside the droplets. If the duration inside or outside a droplet is inferior to $25$ $10^3 \delta t$, we consider it as a false entrance in a droplet and a false exit from a droplet, respectively. This procedure allows to prevent an artifact in the distribution of residence times as particles can oscillate at the border of a droplet.

A similar process is used to compute the the mean time to enter a droplet ($\tau_{\rm ent}$).

%\bibliography{PRL}

%apsrev4-2.bst 2019-01-14 (MD) hand-edited version of apsrev4-1.bst
%Control: key (0)
%Control: author (8) initials jnrlst
%Control: editor formatted (1) identically to author
%Control: production of article title (0) allowed
%Control: page (0) single
%Control: year (1) truncated
%Control: production of eprint (0) enabled
%

\end{document}

\onecolumngrid

\beginsupplement

\begin{center}
	
	{\bf \large Control of encounter kinetics by chemically active droplets}
	
	\textbf{\textit{\large Supplementary Information}}

	$ \ $

Jacques Fries, Roxanne Berthin, Marie Jardat, Pierre Illien, Vincent Dahirel
	
	\textit{Sorbonne Universit\'e, CNRS, Physicochimie des \'Electrolytes et\\ Nanosyst\`emes Interfaciaux (PHENIX), Paris, France}

\end{center}

%\tableofcontents

\section{Size distribution and stability analysis}

To identify the droplets at each timestep, a cluster analysis of the trajectories is performed. Every  $ n\ \delta t$, the positions of the $B$ particles (and $B'$ in the presence of clients) are stored. The distances between these particles are computed. Two particles are assumed to be in the same droplet if they are at a distance $d \le 1.5\sigma$. This value corresponds to the typical distance between $B$ particles in a droplet~\cite{berthin2025}. The volume $v$ of a droplet is assumed to be the sum of the volume of each particle of the droplet. This yields for $\mathcal{N}$ particles : $v=\mathcal{N}\cdot\frac{4}{3}\pi(\frac{\sigma}{2})^3$. 

\begin{figure}[b] 
    \centering
    \includegraphics[width=10cm]{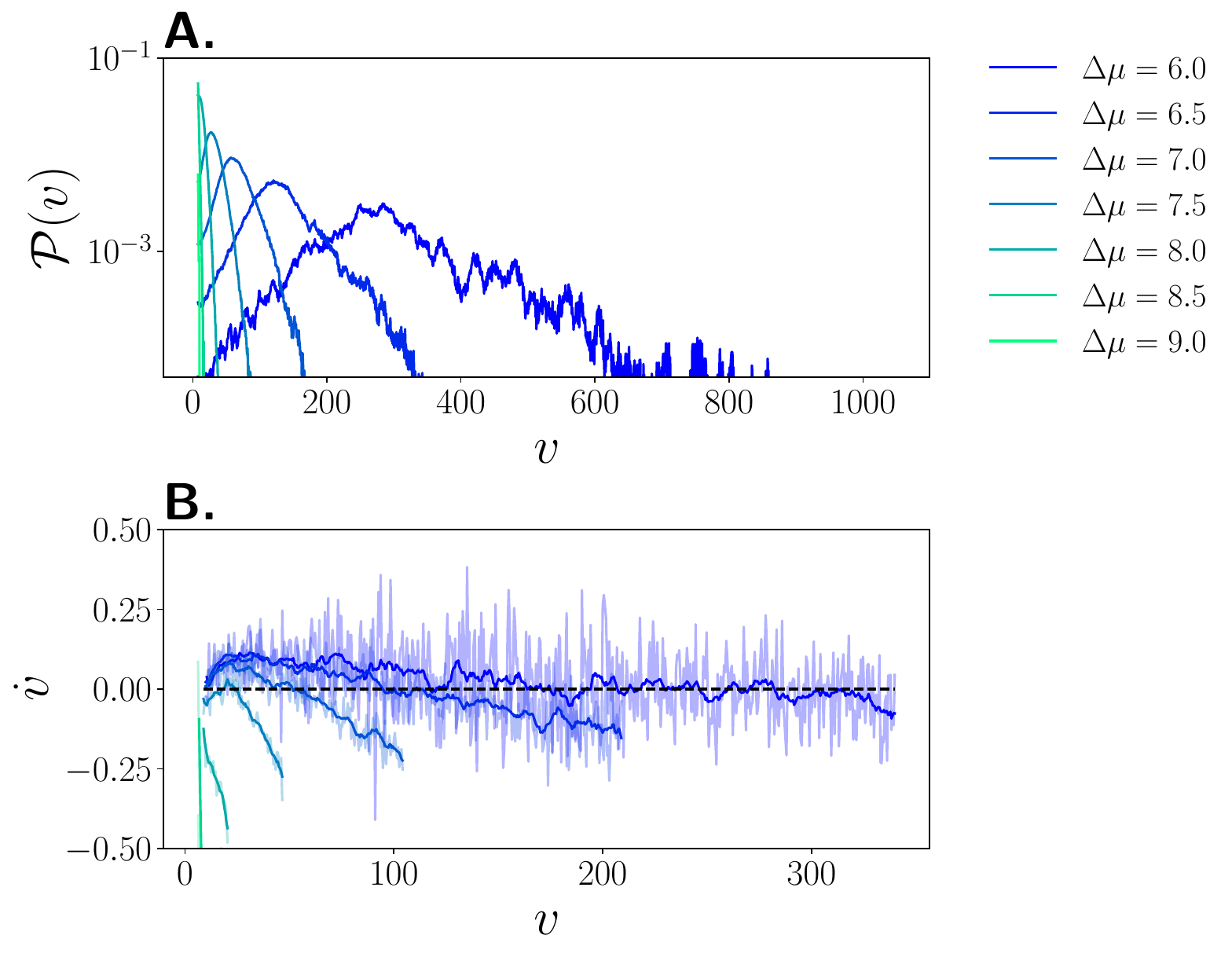}
    \caption{\textbf{A. Probability distribution of droplet volume $\mathbf{v}$ at stationary state} for various values of the chemical drive $\dmu$. \textbf{B. Phase portrait of the volume of droplets }(derivative of the volume with respect to time as a function of the volume) for different values of the chemical drive $\dmu$.}
    \label{fig:supp_droplet_size_selection}
\end{figure}

From this cluster analysis, the probability distribution of droplet number and droplet volume can be derived (see example in Fig.~\ref{fig:supp_droplet_size_selection}.A.). By tracking the droplets over time~\cite{berthin2025}, we can compute the average derivative of the volume $v$ with respect to time as a function of $v$ (referred to as the phase portrait of the system; see example in Fig.~\ref{fig:supp_droplet_size_selection}.B.). Both size distribution and phase portrait highlight the droplet size selection: There is a local maximum of the size distribution, which corresponds to a stable fixed point of the phase portrait~\cite{berthin2025}. We are interested in stable droplets, which corresponds to droplets bigger than the nucleation volume, the latter corresponding to the unstable fixed point of the phase portrait. Thus, only droplets made of at least 35 particles were considered.

\section{Kinetics of encounter between Brownian particles: survival probability}

This paragraph completes the method section from the main text on the estimation of the mean encounter time between two particles in the Brownian Dynamics simulations. 
At each timestep $\delta t$, the distance $d$ between each neighboring particle is computed. The encounter happens when $d \le \sigma$. We compute the fraction of pairs that have encountered before $t$. This fraction is the ratio of the number of pairs of particles having encountered at least once over the total number of pairs. The time evolution of this ratio yield the cumulative distribution function of encounter $F(t)$. 

\begin{figure}
    \centering
    \includegraphics[width=10cm]{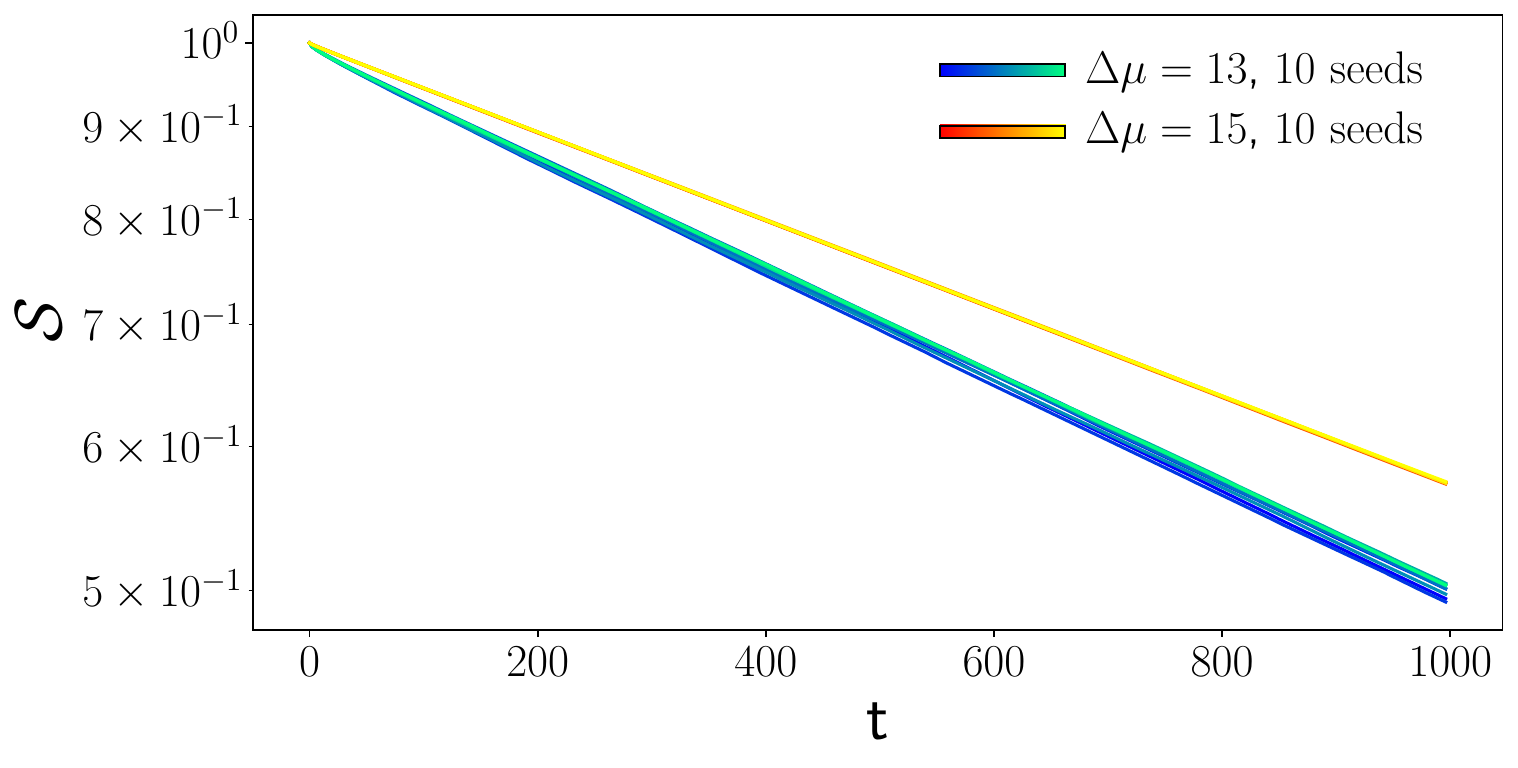}
    \caption{\textbf{Survival probability of encounter between pair of droplet materials (in $A$ or $B$ state)} computed at stationary state for 10 repetitions of each systems (different seeds) and various chemical drive $\dmu$}
    \label{fig:supp_survival_probability}
\end{figure}

We represent in Fig.~\ref{fig:supp_survival_probability} the survival probability $S=1-F$. 
According to our mean field model (see main text), under some assumptions, the survival probability is a linear combination of decreasing exponential functions. Each exponential decrease is associated with a characteristic time. If the times are distinct enough, they can be extracted from a semi-log representation of $S$. 
We find mono-exponential behavior at long time in all cases, and thus define a mean encounter time as the characteristic time associated with the long time exponential decay of $S$.

\section{Diffusion coefficients of droplet particles}

\begin{figure}
    \centering
    \includegraphics[width=10cm]{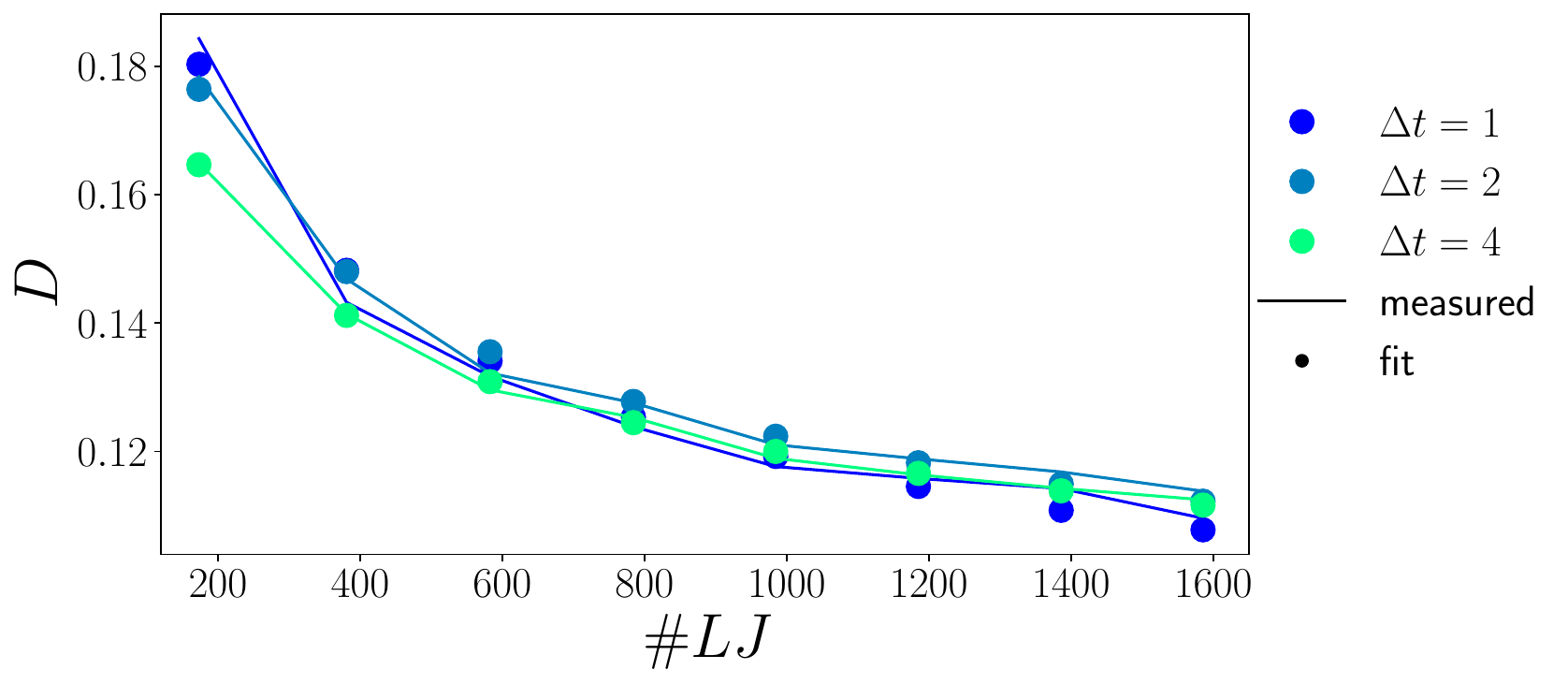}
    \caption{Diffusion coefficient of LJ particles in droplets, as a function of the number of LJ particles in the system.}
    \label{fig:diff_coeff}
\end{figure}

This paragraph completes the method section from the main text on the estimation of the diffusion coefficients of droplet particles in the Brownian Dynamics simulations. 
Instead of computing the Mean Square Displacement (MSD) as a function of t, we compute the averaged MSD of a given particle $i$ for subparts of the trajectories of duration $\Delta t$, where the particle $i$ does not leave the droplet. 
For systems at equilibrium, we measure such MSD inside the droplets for $\Delta t= 1,2,4$. 
We deduce from the MSD the diffusion coefficient of the systems represented in Fig.~\ref{fig:diff_coeff}.

\section{Continuous description of active emulsions}

In order to better explore the coupling between the chemical drive $\dmu$ and a flow at the interface between the droplet and the continuous phase, we resort to a field representation of chemically active droplets. The model is coarse-grained so as to write down a deterministic diffusion-reaction equation of a ternary mixture described by Flory-Huggins free energy and mean-field chemical kinetics~\cite{Zwicker2022}. %Active reactions are associated with similar parameters $\dw$ and $\dmu$. 

\subsection{Formalism}

We consider a $2$-dimensional ternary mixture with species $\{A,B,S\}$. %, where species $B$ phase separate. 
We choose a Flory-Huggins~\cite{floryhuggins} free energy density of the system such that
\begin{align}
    \frac{\lambda^2 f(\phi_A,\phi_B)}{\kB T}&=\sum_{i\in\{A,B\}} \phi_i \ln{(\phi_i)} + (1-\phi_A-\phi_B)\ln{(1-\phi_A-\phi_B)} \nonumber \\
    &+\sum_{i\in\{A,B\}} w_i \phi_i+\ \sum_{i,j\in\{A,B\}} \frac{\chi_{ij}}{2} \phi_i \phi_j - \frac{\kappa_{ij}}{2} \Vec{\nabla}\phi_i \cdot \Vec{\nabla}\phi_j,
    \label{eq_density_free_energy_ternary}
\end{align}
where $(w_i)_{i\in \{A,B\}}$ are the internal energies, $(\chi_{ij})_{i\in \{A,B\}}$ and $(\kappa_{ij})_{i\in \{A,B\}}$ are interaction parameters. Their expressions verify  
\begin{align}
    \chi_{ij}^*&=2 \frac{e_{ij}+e_{ss}-e_{is}-e_{js}}{\kB T},\\
    \kappa_{ij}^*&=\lambda^2 \frac{e_{ij}+e_{ss}-e_{is}-e_{js}}{2\kB T},
\end{align}
where $(e_{ij})_{i\in \{A,B,S\}}$ is the interaction energy matrix between each species and $\lambda$ is the typical radius of a particle. We choose $e_{ij}=0$ if $i,j \ne B,B$ such that Eq.~\ref{eq_density_free_energy_ternary} becomes 
\begin{align}
    \frac{\lambda^2 f(\phi_A,\phi_B)}{\kB T}&=\sum_{i\in\{A,B\}} \phi_i \ln{(\phi_i)} + (1-\phi_A-\phi_B)\ln{(1-\phi_A-\phi_B)} \nonumber \\
    &+\sum_{i\in\{A,B\}} w_i \phi_i+ \frac{\chi_{BB}}{2} \phi_B^2 - \frac{\kappa_{BB}}{2} \Vec{\nabla}\phi_B \cdot \Vec{\nabla}\phi_B.
    \label{eq_density_free_energy_ternary2}
\end{align}
The chemical potentials read
\begin{align}
    \frac{\mu_A}{\kB T} &= \ln{(\phi_A)} - \ln{(1-\phi_A -\phi_B)}+ w_A\\
    \frac{\mu_B}{\kB T} &= \ln{(\phi_B)} - \ln{(1-\phi_A -\phi_B)}+ w_B  + \bigg(\chi_{BB}\phi_B + \kappa_{BB} \nabla^2 \phi_B\bigg)
\end{align}

We consider passive and active reactions interconverting $A$ and $B$, respectively written $A \rightleftharpoons B$ and $B + {\rm ATP} \rightleftharpoons A + {\rm ADP} $. The detail balance of the rates~\cite{Weber2019} implies that the passive and active reaction fluxes $s_p$ and $s_a$ verify 
\begin{align}
    \frac{s_p^\rightarrow}{s_p^\leftarrow}&=\exp{(-\beta (\mu_B-\mu_A))},\\
    \frac{s_a^\rightarrow}{s_a^\leftarrow}&=\exp{(-\beta (\mu_B+\mu_{\rm ATP}-\mu_A-\mu_{\rm ADP})}).
\end{align}

The chemical drive with this formalism corresponds to the free energy of ATP hydrolysis into ADP, $ \dmu = \mu_{\rm ATP}-\mu_{\rm ADP}$, which is fixed by a chemostat. 

Because the conversion of B into A is associated with the favorable ATP hydrolysis, the active reaction promotes this sens of reaction. Moreover, in order to promote the destruction of  B particles in the droplets (as proposed by Zwicker et al. as a mechanism to control droplet size~\cite{Zwicker2022}), the kinetic rates are modified accordingly: the passive rate is multiplied by  $(1-\phi_B)$, while the active one is multiplied by $\phi_B$. Therefore, the equations describing the evolution of the fields are
\begin{align}
        \frac{\partial\phi_A}{\partial t}&=\lambda_{A}\Vec{\nabla}\cdot(\Vec{\nabla} \mu_A)+ k_p(1-\phi_B) (e^{\beta\mu_B}-e^{\beta\mu_A}) \nonumber\\
    &+k_a \phi_B (e^{\beta(\mu_B+\mu_{\rm ATP})}-e^{\beta(\mu_A+\mu_{\rm ADP})})\\
        \frac{\partial\phi_B}{\partial t}&=\lambda_{B}\Vec{\nabla}\cdot(\Vec{\nabla} \mu_B)- k_p(1-\phi_B) (e^{\beta\mu_B}-e^{\beta\mu_A}) \nonumber\\
    &-k_a \phi_B (e^{\beta(\mu_B+\mu_{\rm ATP})}-e^{\beta(\mu_A+\mu_{\rm ADP})})
    \label{eq_field_evolution_ternary}
\end{align}
where $\lambda_{A}$ and $\lambda_{B}$ are mobilities. The partial differential equations are solved on a $128 \times 128$ periodic grid using the python package Py-pde~\cite{zwicker_py-pde_2020}.
The initial values of the density fields are chosen from a uniform distribution, such that $\phi_A \in [0.18,0.22] $ and $\phi_B \in [0.48,0.52]$.

% \begin{figure}
%     \centering
%     \includegraphics[width=\linewidth]{new_figs/supp_encounter_time_in_fct_t_in_t_smol.png}
%     \caption{\textbf{Characteritic time of encounter in function of the average time spent by $\mathbf{B}$ particles inside droplets divided by the typical time for two particles to encounter.} The quantity are computed at stationary state for systems yielding droplets with selected size with different internal energy $\dw$ and chemical drive $\dmu$. The two quantity have the same probability distribution with p-value$=0.062$ calcultated with two-sided Kolmogorov-Smirnov test.}
%     \label{fig:supp_encounter_time_in_fct_t_in_t_smol}
% \end{figure}
%\bibliography{/Users/pierreillien/work/docs/library.bib}

\subsection{Results}

\begin{figure}
    \centering
    \includegraphics[width=0.5\linewidth]{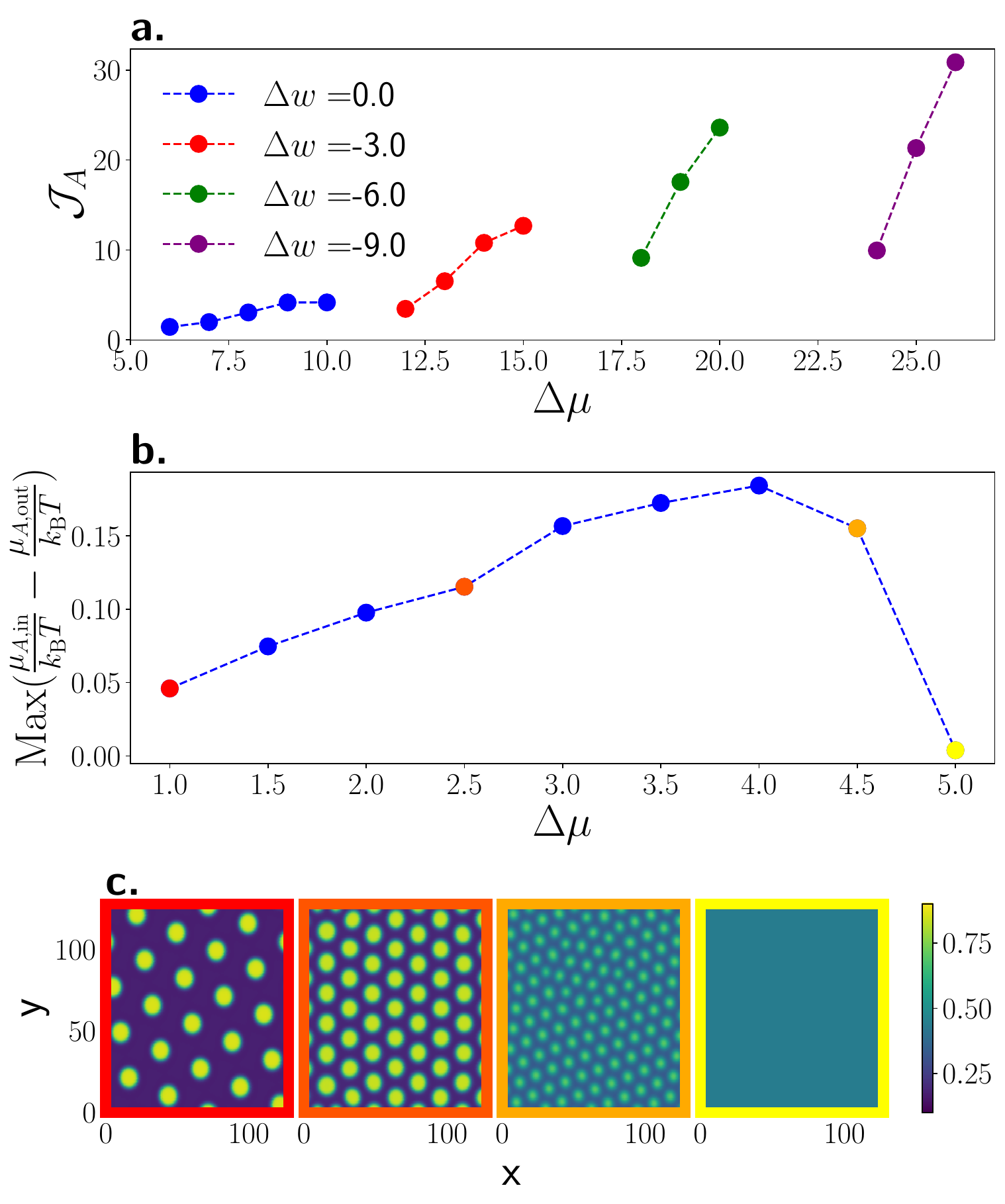}
    \caption{\textbf{(a)} Net rate of $A$ escape from droplets $\mathcal{J}_A = (-dN_A/dt)_{\rm diff}$, computed from BD simulations.
    \textbf{(b)} Maximal difference of steady state chemical potential of $A$ species inside and outside droplets in continuous systems.
    The partial differential equations are solved %SUPP on a $128 \times 128$ grid 
    using the py-pde package~\cite{zwicker_py-pde_2020}, with these parameters : $w_B=0$, $\chi=2.5$, $\kappa=1$, $\lambda_A=\lambda_B=1$, $k_{\rm passif}=k_{\rm actif}=0.01$, $\mu_{\rm ATP}=0$. $w_A$ and $\mu_{\rm ADP}$ are varied to modify $\dmu$ (with $\dmu=\mu_{\rm ADP}-\mu_{\rm ATP}$) and $\dw$ (with $\dw=w_B-w_A$). \textbf{(c)} The density of continuous systems at stationary state is given in the snapshots  in the bottom figures for several systems. $B$-rich regions are colored in yellow. The frame color matches the color of the corresponding point in \textbf{(b)}.}
    \label{fig:chemical_potential_and_flux}
\end{figure}

Using this approach, we evaluate how the chemical drive leads to localized chemical potential gradients of $A$ or $B$ particles at steady state.  
Fig.~\ref{fig:chemical_potential_and_flux}-(b) shows the evolution of the maximal difference of chemical potential of the $A$ species inside and outside a droplet as a function of the chemical drive $\dmu$. 
The chemical potential gradient increases for increasing values of $\dmu$, unless the chemical drive is so strong that droplets are unstable ($\dmu > \dmu_{\rm crit}$). 
Stronger gradients lead to stronger flows of particles.

These results confirm our findings shown in the main text: chemical energy can be efficiently converted into mechanical work leading to collective flows of particles at the interface between the droplets and the continuous phase. 

%\bibliography{biblio_supp}

%merlin.mbs apsrev4-1.bst 2010-07-25 4.21a (PWD, AO, DPC) hacked
%Control: key (0)
%Control: author (8) initials jnrlst
%Control: editor formatted (1) identically to author
%Control: production of article title (-1) disabled
%Control: page (0) single
%Control: year (1) truncated
%Control: production of eprint (0) enabled
%